\documentclass[nohyper, letterpaper,11pt,notoc]{JHEP3}
\usepackage{amsmath,amssymb,amscd,graphicx,xy,epsfig}
\newcommand{\ba}{\begin{eqnarray}}
\newcommand{\ea}{\end{eqnarray}}
\newcommand{\be}{\begin{equation}}
\newcommand{\ee}{\end{equation}}

\newcommand{\cL}{{\cal L}}

\newcommand{\MSSM}{{\mathsf{MSSM}}}
\newcommand{\HIDDEN}{{\mathsf{HIDDEN}}}

\newcommand{\GeV}{~\mathrm{GeV}}
\newcommand{\TeV}{~\mathrm{TeV}}

\newcommand{\is}{\!&\! = \! & \!}

\newcommand{\newsection}[1]{
\addtocounter{section}{1} 
\setcounter{subsection}{0} \addcontentsline{toc}{section}{\protect
\numberline{\arabic{section}}{{\rm #1}}} \vglue .0cm \pagebreak[3]
\noindent{\large \bf  \thesection. #1}\nopagebreak[4]\par\vskip .1cm}
\newcommand{\newsubsection}[1]{
\addtocounter{subsection}{1}
\addcontentsline{toc}{subsection}{\protect
\numberline{\arabic{section}.\arabic{subsection}}{ #1}} \vglue .0cm
\pagebreak[3] \noindent{\it \thesubsection. #1}\nopagebreak[4]\par\vskip .1cm}

\newcommand{\aalpha}{{\mathsf A}}

\newcommand{\bbeta}{{\mathsf B}}
\newcommand{\hstar}{\raisebox{-2.5pt}{*}}

\newcommand{\hhstar}{\raisebox{-2.5pt}{*}}
\newcommand{\wwedge}{\hspace{1pt} }
\newcommand{\wwwedge}{\hspace{-1pt}\wedge\hspace{-1pt}}

\newcommand{\cA}{{\mathcal A}}
\newcommand{\cF}{{\mathcal F}}

\newcommand{\del}{\partial}

\newcommand{\spc}{\hspace{1pt}}

\title{A Higher Form (of) Mediation}
\author{Herman Verlinde,${}^a$ Lian-Tao Wang,${}^a$ 

Martijn Wijnholt,${}^b$ and Itay Yavin\,${}^a$ \\

\it ${}^a${ Physics Department,
 Princeton University,
 Joseph Henry Laboratories, Princeton NJ 08540}\\
${}^b$ {\it Max-Planck-Institut f\"ur Gravitationsphysik,
Albert-Einstein-Institut, 
Potsdam, Germany}}

\abstract{We exhibit a simple and robust mechanism for bulk mediation of supersymmetry
breaking between hidden and visible sectors localized on 
geometrically separated
D-branes in type II string theory. The mediation proceeds via RR p-forms
that couple via linear Chern-Simons terms to the abelian vector bosons on the branes.
From a 4-d low energy perspective, the mechanism reduces to $U(1)$ mediation.
}

\begin{document}
\renewcommand{\footnotesize}{\small}

\addtolength{\baselineskip}{.9mm}

{\hbox to\hsize{ \hfill PUPT-2251}}

%
%
%

\renewcommand{\footnotesize}{\small}

\newcommand{\sumi}{\mbox{\large $\sum\limits_{\raisebox{.5mm}{\scriptsize $i$}}$}}
\newcommand{\QQ}{\mbox{\hspace{1pt}$q$}}
\newcommand{\YY}{\mbox{\small $Y$}}
\addtolength{\abovedisplayskip}{1mm}
\addtolength{\belowdisplayskip}{1mm}

\addtolength{\abovedisplayshortskip}{1mm}
\addtolength{\belowdisplayshortskip}{1mm}

\newcommand{\heart}{\raisebox{-1.5pt}{\large $\heartsuit$}}

\newcommand{\diamant}{\mbox{\large $\diamondsuit$}}

\newcommand{\oomega}{{\mbox{\large $\omega$}}}

\newcommand{\avg}[1]{\left\langle #1 \right\rangle}

\newsection{Introduction}

A key component in string phenomenology is to find a good mechanism of
supersymmetry breaking, and for mediating its effect to the standard
model. In typical phenomenological scenarios, SUSY breaking takes
place in a hidden sector 
of particles without direct couplings to the standard model
particles. The two sectors  communicate via messenger particles, that
mediate the SUSY breaking effects and thereby induce a distinctive
pattern of MSSM soft terms. While there are several promising
phenomenological scenarios of mediated SUSY breaking, 
there has been only limited success in finding robust realizations in
string theory. 

A logical setting for scenarios with
messenger mediated supersymmetry breaking is via
string models in which the standard model is localized on branes
\cite{Dreview}\cite{bottomup}\cite{bjl}. 
The decoupling of the hidden and visible sector is then achieved via geometric separation
in the extra dimensions. A typical set-up is indicated in fig 1. It consists of a visible brane 
configuration localized in some region $\heart$, that we assume realizes the MSSM, or some viable
extension thereof.  We assume that supersymmetry is broken on the hidden brane located at
$\diamant$.  In this paper we will exhibit, in this general set-up,
a simple and robust new mechanism for mediating SUSY breaking to the MSSM sector.

In the set-up of fig 1, the hidden brane communicates with the
MSSM brane via closed string fields -- graviton, dilaton and p-form fields --
that traverse the Calabi-Yau geometry. Which particular bulk field dominates the
mediation depends on specific details, such as moduli stabilization,
presence of warping, and other properties of the compactification geometry 
\cite{Anisimov:2002az}\cite{Choi:2005ge}\cite{Diaconescu:2005pc}\cite{Kachru:2007xp}.
This sensitivity on details of the Planck scale geometry is a
precarious aspect of all bulk mediation mechanisms, that typically makes
it hard to extract robust, universal predictions. 
A concrete danger is that the soft terms may induce new CP or flavor 
violations, that would conflict with current experimental bounds. 
An attractive Planck scale scenario, that manages to avoid these
problems, is anomaly mediation  
\cite{anomaly}.  It relies on the assumption that  all bulk modes
except gravity are massive compared to the distance between the hidden and
visible brane. The mediation then proceeds via the superconformal anomaly. In its minimal form, however,  it predicts a negative mass squared for the sleptons.
Recent studies that investigate string realizations of anomaly mediation include
\cite{Anisimov:2002az}
\cite{Choi:2005ge}\cite{Kachru:2007xp}.

Another promising mechanism for communicating SUSY breaking is 
gauge mediation \cite{gauge}.  In this class of models, the SUSY
breaking scale can 
be much lower. The  messenger fields are chiral multiplets,  that
couple to the hidden  
sector and carry standard model gauge charges. 
Flavor violations are suppressed, because the messengers couple to the
visible sfermions via 
universal gauge interactions. In brane realizations, gauge mediation
proceeds via open 
strings stretched between the visible and hidden sector branes. The
branes must therefore be 
placed at a small sub-stringy relative distance  \cite{Diaconescu:2005pc}.

\begin{figure}[t]
 \begin{center}
 \includegraphics[width=3.9in]{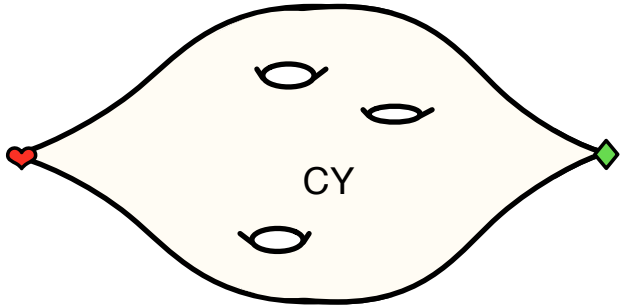} 
\caption{In our set-up, the MSSM and the SUSY breaking hidden sector
each live on collections of branes localized within
separate small regions of the Calabi-Yau manifold.}
\end{center}
\end{figure}

\newcommand{\ccc}{\mbox{\small $\cal C$}}
\newcommand{\four}{{\mbox{\tiny $4$}}}
\newcommand{\six}{{\mbox{\tiny $6$}}}
\newcommand{\cS}{{\mbox{\small $\mathsf{S}$}}}

Based on the available scenarios, one may conclude that there are two basic
possibilities:

\smallskip

 \noindent
$i)$ The separation $d$ between the branes is larger than the string length
\be
 \hspace{2cm} d>\!\! > \ell_s \hspace{1cm}
\ee
and the supersymmetry breaking takes place via Planck scale mediation, effectuated by
closed string exchange.  Soft masses
are of order $\avg{F}/M_{Pl}$ or smaller and the SUSY breaking scale must be higher than the geometric mean
of the electro-weak and Planck scale.
\smallskip 
 
\noindent
 $ii)$ The hidden and visible branes are at sub-stringy relative distance
\be
\noindent
\hspace{2cm} d <\!\!< \ell_s \, .\hspace{1cm}
\ee
SUSY breaking then takes place via gauge mediation, brought about by
light open strings of mass $M= d/\ell_s^2$. The SUSY breaking scale
can now be much smaller than the Planck scale. The soft masses are of
order $\avg{F}/M$ with $M$ the messenger scale. 

\smallskip

In this paper we will uncover a third possibility: we will exhibit a
simple mechanism in which both the closed and {open} string modes
participate in the communication of supersymmetry breaking. Moreover, 
the mechanism works irrespective of the brane separation,
and both {low} and {high scale} supersymmetry breaking
are allowed. This new mediation scenario 
manifestly respects flavor symmetry and relies on the coupling between
$U(1)$ vector bosons on the branes and the 
 RR-forms in the bulk.  In a suitable set-up, it can dominate over all
 other mediation mechanisms.

\bigskip

\newsubsection{U(1) Mediated Supersymmetry Breaking}

Phenomenological string models frequently lead to standard model-like
gauge theories 
with additional $U(1)$ factors. In case both the visible and
hidden sector particles are charged under a common $U(1)$ gauge group, the associated
$Z'$-boson can be responsible for the mediation of supersymmetry
breaking.  The phenomenology of such $U(1)$ mediation scenarios, in which the supersymmetry breaking is parametrized by the gaugino mass, 
was recently considered in \cite{LPWY}.
The low energy action takes
the schematic form
\be
\label{vihi}
 \cL = \cL_{\MSSM}(Q, \cA^\prime) +\;  \cL_{\HIDDEN}(X , \cA^\prime)\, .
\ee
Here 
$Q$ and $X$  denote the MSSM and hidden sector fields and $\cA'$ the
$U(1)$ gauge field. 
In the hidden sector, supersymmetry is broken by some F-term VEV
$\langle F\rangle$. On the visible brane, SUSY breaking is driven by
the  mass splitting between the $\cA'$ boson  and its
superpartner~$\lambda^\prime$.

In its minimal form, $U(1)$ mediation realizes an intermediate version
of split supersymmetry. 
All visible matter multiplets that are
 charged under the $U(1)$ receive soft mass terms at 1-loop, while
 the gauginos, which do not directly couple to $\cA^\prime$, acquire a
 mass at the 2-loop level. The observed lower 
bound on the gaugino mass (of about $100\GeV$) thus forces the sfermion mass
to be  of order $100 \TeV$. 
To generate the electroweak scale, one must accept a finetuning of order $10^{-6}$. 
A more detailed discussion of the phenomenology of $U(1)$ mediation is
found in \cite{LPWY}.

In this note, we point out how $U(1)$ mediation is naturally realized in string theory.
Having such a embedding at hand is of particular interest if it
allows for direct investigation of the required tuning mechanism via the multitude of vacua in
the string landscape. For this reason, we will imagine a set-up as in fig 1, with the hidden and
visible sector on branes located at local regions of the Calabi-Yau manifold, each at the
bottom of a flux stabilized warped throat \cite{KS}\cite{GKP}. The discrete tuning of the flux then provides 
explicit control over the relative ratio between the string scale and the low energy scale
on the brane.

At first sight, the set-up of fig 1 looks like the wrong starting point for $U(1)$ mediation: since both
branes are separated by a large  distance compared the string scale, they do not share
any low energy open string modes. The two branes each have their own world-volume gauge theory,
with each their own $U(1)$ vector boson, $\cA_V$ and $\cA_H$. The open string action  therefore
 splits up as (here $Q$ and $X$ encode all other visible and hidden sector fields) 
\ba
\label{open}
\cL_\heartsuit + \cL_\diamondsuit = \cL_{\MSSM}(Q, \cA_V) + \cL_{\HIDDEN}(X ,\cA_H). 
\ea
This is clearly not a suitable action for any mediation mechanism between the branes.
However, the action (\ref{open}) does not tell the complete story,
since it ignores the coupling to closed string modes that propagate in the bulk region between the branes.

In the following we will show how, via a simple topological arrangement,
the bulk physics can enforce a  low energy identification between $\cA_V$ and $\cA_H$.
The basic mechanism is well known and makes use of the linear Chern-Simons-coupling of  the gauge fields on the branes to the
RR p-form field in the bulk. 
In a suitable geometrical set-up, this RR p-form
field leads, upon KK reduction, to a massless 2-form field $\ccc$ with a 
4-d action of the form\footnote{This interaction between abelian gauge bosons and 2-form fields  is familiar as part of the Green-Schwarz anomaly cancelation mechanism \cite{GS} for anomalous $U(1)$'s, and results in a mass term for the corresponding vector boson \cite{ALE}\cite{Ghilencea:2002da}. However, as is well known (and will be explained later),  the RR-form mechanism for generating $U(1)$ masses applies equally well to non-anomalous $U(1)$s \cite{Dreview}\cite{jlouis}\cite{buican}. We will focus on the case where the $U(1)$ is non-anomalous. } 
\be
\label{een}
\cL_{RR} \; = \; \ccc \wwwedge( d \cA_V +  d \cA_H)
\; + \; 
{1\over 2 \mu^2} \; |\spc d \spc \ccc\spc |^2 \, .
\ee
This action 
can be recognized as  a mass term for the gauge field $\cA_+ = \cA_V + \cA_H$. The simplest way to see this, is to
rewrite $\Delta \cL$ in the dual form
\be
\label{twee}
\cL_{RR} \; = \; (\cA_V + \cA_H) \wwwedge H 
\; + \; 
{1\over 2 \mu^2} \; |\spc H \spc |^2\; + \; d\varphi \wwwedge H 
\ee
where $H$ and $\varphi$ are independent fields. This action is gauge invariant, provided that the 
field $\varphi$ shifts under the $U(1)$ gauge rotations. The equivalence of (\ref{twee}) to (\ref{een}) is evident: 
the equation of motion of $\varphi$ reads $dH=0$, which can be solved by setting $H=d\spc \ccc$.
However, instead of solving for $H$, one may also integrate out $H$ and obtain
\be
\label{drie}
\cL_{RR} \, = \, {1 \over 2} \mu^2 |d\varphi +\cA_H +\cA_V|^2\, .
\ee
This is the well-known string mechanism for generating St\"uckelberg mass terms
for abelian vector bosons.
The mass scale $\mu$ is set by the size of the compactification manifold; for string scale 
size compactifications, $\mu$ is of order the string scale. 

\begin{figure}[hbtp]
 \begin{center}
\includegraphics[width=3.8in]{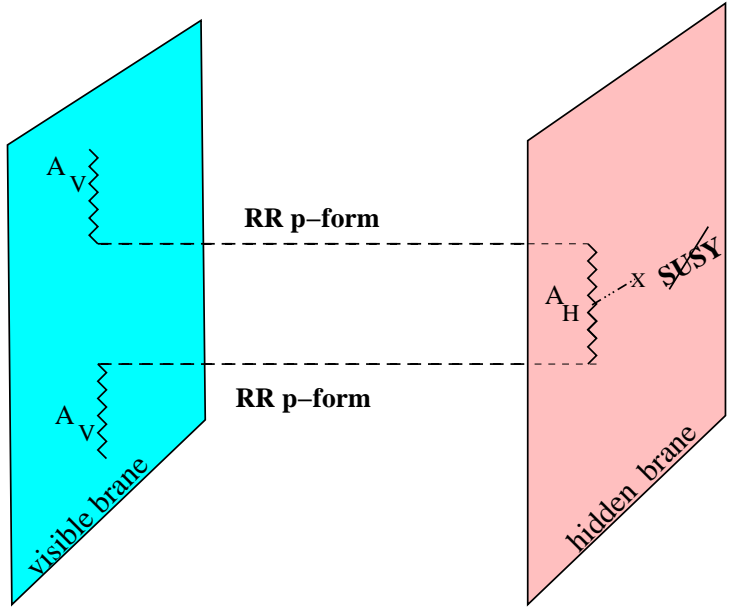} 
\caption{SUSY breaking on the hidden brane gets mediated to the visible sector
via an RR p-form. It manifests itself via a mass-splitting between the abelian vector boson
$\cA_1$ and its superpartner. 
}
\end{center}
\end{figure}

The upshot of the story is that, via the interaction between the branes and the RR-bulk fields,  
the linear sum $\cA_+ = \cA_V + \cA_H$ effectively decouples from the low energy physics, leaving
a low energy action of the form (\ref{vihi}) with 
\ba
\cA^\prime = \cA_H-\cA_V.
\ea
Evidently, assuming that the hidden brane hosts a SUSY breaking sector, the system then provides a string realization of $U(1)$ mediation. Somewhat unexpectedly, 
it arises as a bulk mediation mechanism. Pictorially, we can think of the mechanism as  depicted in fig 1: the vector boson $\cA_1$ on the visible brane can, via the linear CS-coupling, convert itself into an RR p-form that traverses the bulk towards
the hidden brane. Here it can pick up a SUSY-breaking contribution, which via the same RR p-form
mechanism, can be transported back to the visible sector. It is important to note that the mechanism
only works for abelian vector bosons, since the RR p-forms do not have a linear CS-coupling to 
non-abelian gauge bosons.

This paper is organized as follows. In section 2, after a short review the basic building blocks  of string phenomenology with D-branes, we present some more details of the RR-form mediation mechanism.
In section 3 we discuss some possible applications of our result. In the Appendix, we summarize
 how $U(1)$ mediation can arise in the heterotic string.

\bigskip
\bigskip


\newsection{RR-form Mediation}

In this section we will display the RR-form mediation mechanism in a specific
IIB string theory set-up. Although the main ingredients and arguments 
are all quite familiar to experts, the story has sufficiently many 
instructive aspects 
to warrant a more detailed exposition. 

Our set-up is as in fig 2: we imagine that the MSSM and the SUSY breaking hidden sector each live on separate 
(collections of) branes localized within separate small regions of the Calabi-Yau manifold. We will
demonstrate how, in this generic set-up,  the bulk physics may enforce a low energy identification between two abelian vector bosons $\cA_V$ and $\cA_H$ that live on the two separate 
branes. The key step will be to arrange things such that the 
gauge field that gets a mass is a linear sum of the two gauge fields. 
To set the stage, we first summarize some relevant ingredients of local brane constructions of
4-d gauge theories.

\bigskip

\newsubsection{Phenomenology with branes}


Engineering of 4-d gauge theories via D-branes has become a quite
well-developed technology.
Enabled by a growing assortment of available ingredients -- brane intersections,
branes at singularities, orientifolds --
a rich landscape of gauge theories has been constructed, including an increasing number
that closely resemble the MSSM \cite{Dreview}\cite{bottomup}\cite{bjl}.
While string phenomenology with D-brane does not readily encompass grand unified models or enforce gauge coupling unification, it has several other advantages. In particular, it is well suited for a bottom up strategy, that aims to construct the MSSM  in terms of open string modes, that are   
hierarchically and geometrically separated from the string scale dynamics that determines
 the shape of the compactification manifold. Such a dynamical decoupling limit
 can be accomplished
 by  localizing the brane configuration inside a small region of the Calabi-Yau, possibly
 at the bottom of a warped throat.

We will  work in type IIB string theory, but a similar story holds for the IIA case. (We will comment on
the IIA case later.)
IIB string theory has Dp-branes with $p$ any odd integer from 1 to 9.  A typical IIB
brane configuration will involve combination of D3, D5 and D7-branes, that
 fill our 3+1 space-time, while their remaining dimensions
wrap internal compact cycles. Chiral matter arises
from open strings that live at the intersection between D7 and D5 branes.\footnote{Note that 
D5 and D7-branes have 2 and 4 internal dimensions, resp. Since (2 + 4) equals
the dimension of the internal CY manifold, their intersection looks like a point inside the 
CY but fills the 3+1-d space-time.}
 Due to the interaction with the local curved geometry, the D-branes will in general form 
 bound states, known as fractional branes \cite{ALE}. Each bound state has D3, D5 and D7 brane
 components, that respectively wrap $0$-, 2- or 4-cycles supported inside the
 local geometry, see fig 4.\footnote{The name fractional brane is chosen, because via
 the rearrangement of constituent branes into bound states combinations,  a single 
 D3-brane on a CY singularity may split into several fractional branes. Even a single D3 
 can thus support a rich spectrum of gauge and chiral matter. Examples of D3 brane
 theories with MSSM like matter have been considered in \cite{bjl}.}

 The world-volume theory on a collection
 of fractional branes  takes the general form of a quiver gauge theory \cite{ALE}:
each fractional brane gives rise to a $U(N)$ gauge factor, with $N$ its multiplicity, 
and every intersection between two branes produces a chiral matter multiplet in the
corresponding bi-fundamental representation.  These rules can be generalized by including  orientifold planes.
 The challenge of open string phenomenology is to find a suitable
Calabi-Yau singularity and collection of fractional branes, such that the world-brane theory
takes the form of the MSSM, or some
sufficiently realistic extension thereof~\cite{Dreview}.

\begin{figure}[hbtp]
\begin{center}

\includegraphics[scale=0.4]{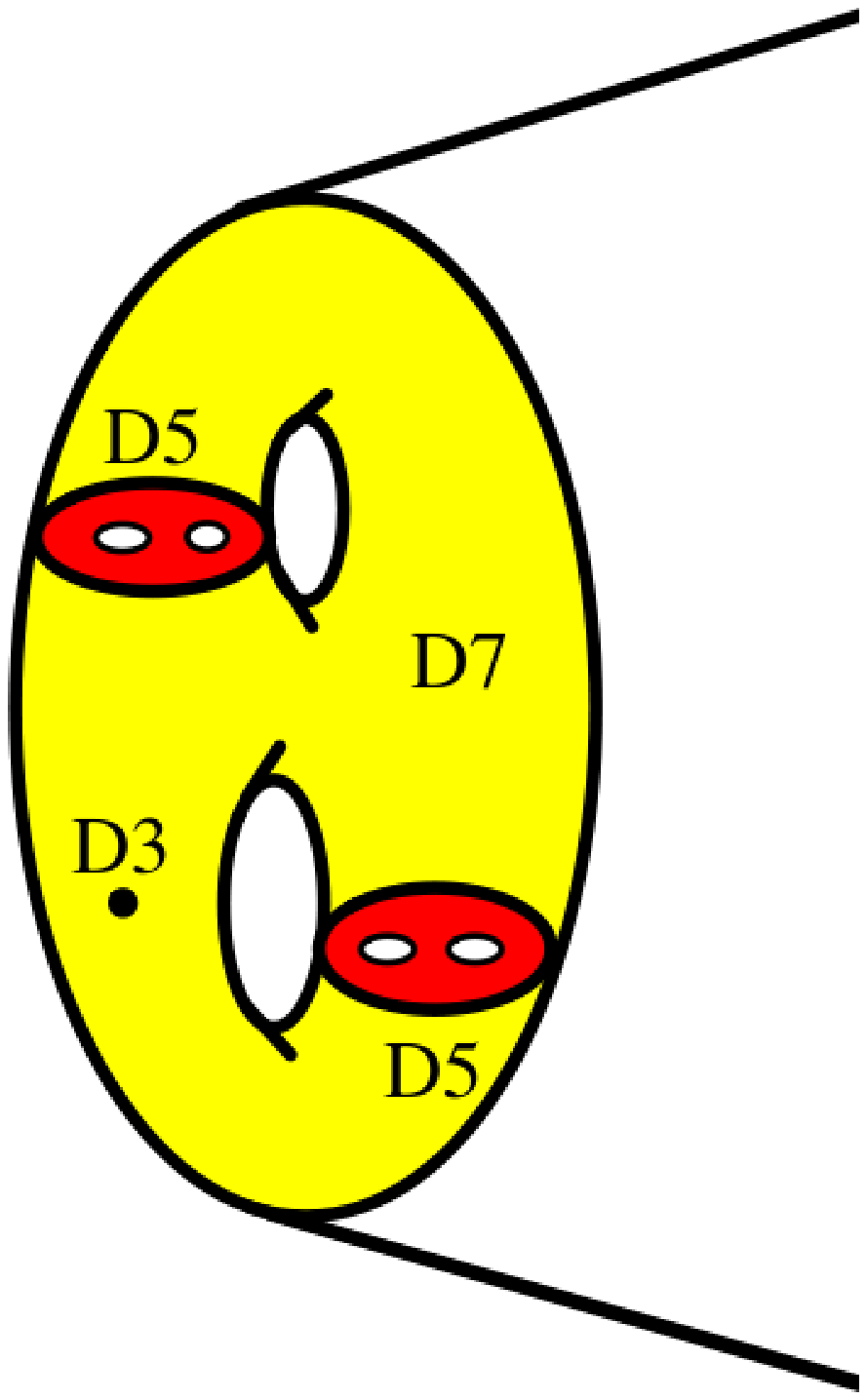} 
\includegraphics[scale=0.34]{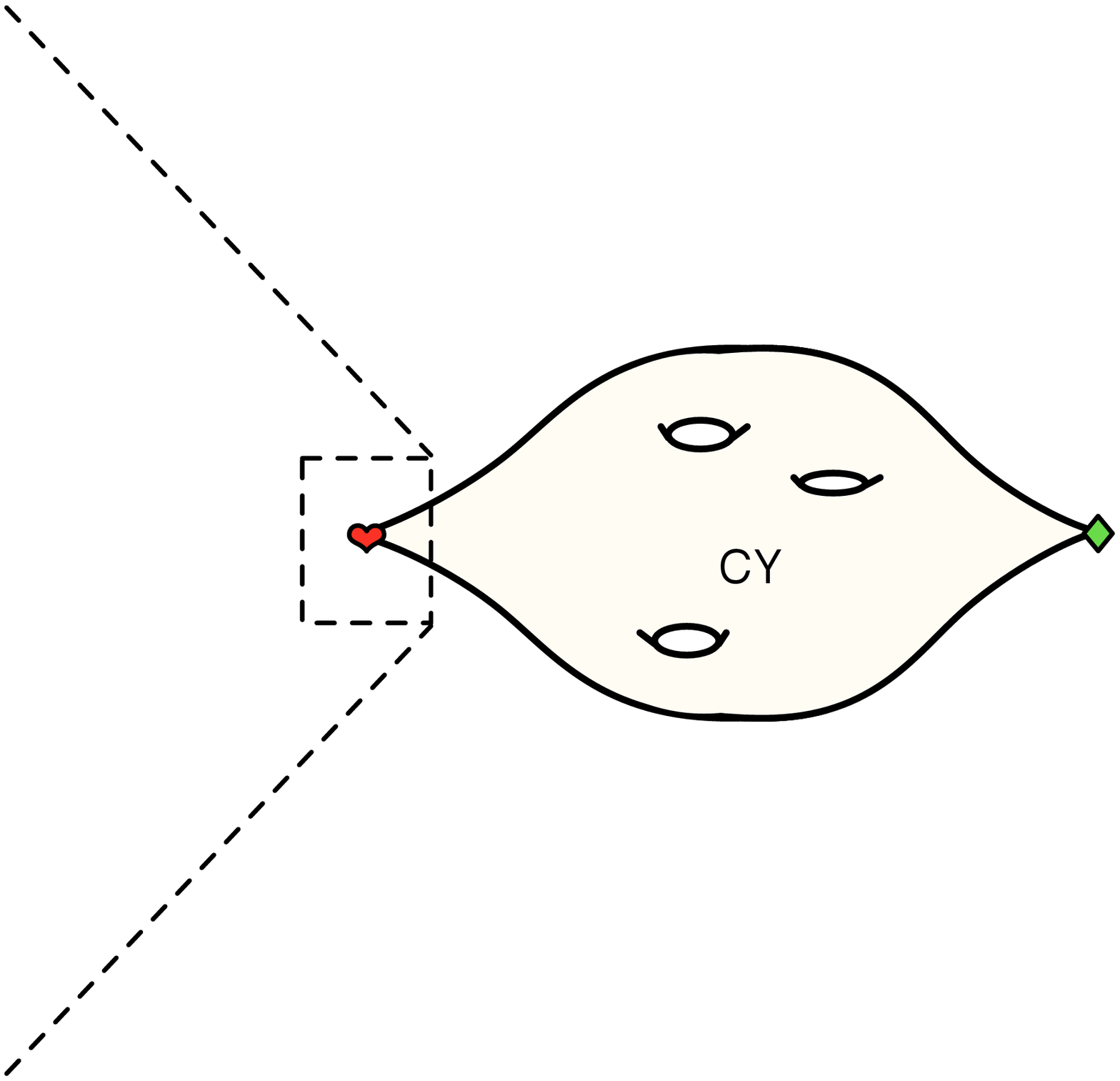} 
\caption{A typical localized IIB  brane configuration consists of bound state combinations of D3, D5 and D7-branes
wrapping 0-, 2- and 4-cycles, supported in a local region of the Calabi-Yau geometry.  The local region may lie at the tip of a warped throat. }
\end{center}
\end{figure}

Most D-brane constructions of MSSM-like theories involve extra $U(1)$ factors besides the
hypercharge symmetry.
The extra $U(1)$ gauge bosons typically acquire a string scale mass via coupling to
closed string RR-form fields. (We will review this mechanism below.)
Which $U(1)$ factors actually survive as low energy gauge symmetries
depends on the topology of how the local CY singularity is embedded inside of the full
compact CY geometry \cite{jlouis}\cite{buican}.  The common occurrence of extra $U(1)$ factors in brane
constructions motivates our present study of their possible role in the mediation
of supersymmetry breaking.

\bigskip
\bigskip

\newsubsection{$U(1)$ Mediation via RR-forms}

Now we turn to the stringy $U(1)$ mediation mechanism.
For concreteness, we assume that the visible $U(1)$ vector boson $\cA_V$ lives on a D5 brane,
that wraps a 2-cycle  $\aalpha_V$ supported within the visible region $\heart$ of the compactification manifold (see figs 1 and~3).
Similarly, the $U(1)$ vector boson $\cA_H$ on the hidden brane formation is assumed to arise from
open strings on a D5 brane,
that wraps a 2-cycle  $\aalpha_H$, supported within the local region $\diamant$.
Note that the D5-branes are
not placed in isolation, but are part of a bound state configuration of branes that hosts all other
visible and hidden matter.  The opens strings supported at the intersection of the D5
and D7-branes in the visible region represent chiral matter charged under $\cA_V$.

 For localized
brane constructions, the K\"ahler area\footnote{This area is defined as $| \int_\aalpha (J + i B)|$ with
$J$ the K\"ahler 2-form and $B$ the NS 2-form field.}
of the 2-cycles $\aalpha_i$ represents the FI parameter
on the the world volume gauge theory. So if this area is finite, it results in a non-zero FI-term\footnote{One intuitive way to quantify this correspondence is to consider a $D3$-brane that wraps the 2-cycle $\aalpha$. In 3+1-dimensions,
it reduces to a D-string with string tension proportional to the area of $\aalpha$, while in the gauge theory
it represents a D-term string, with tension proportional to the FI-term.} that via the D-term constraint
dictates non-zero expectation values for charged matter fields (produced by open string modes located at brane intersections).
 This breaks the $U(1)$ gauge symmetry and renders the abelian gauge boson massive via a conventional Higgs mechanism. 
In the following, however, we will focus on another mechanism for giving a mass to the abelian
vector bosons.

Consider a single isolated $D5$-brane wrapped
on a 2-cycle $\aalpha$. (We will return to our actual set-up with a hidden and visible brane momentarily.) The field theory on the brane is described by a Born-Infeld
 action together with a Chern-Simons term. The CS term contains a linear
coupling 
$$
C_{\it 4} \wwwedge d \cA\, 
$$ 
between the world-volume
gauge field $\cA$ and the RR 4-form field
$C_{\it 4}$. 
The 5-form field strength of $C_{\it 4}$  is self-dual in 10 dimensions
\ba
\label{sd}
F_{\it 5} = \hstar F_{\it 5}\, , \qquad \qquad F_{\it 5} = d C_{\it 4}\, .
\ea 
Although $C_{\it 4}$ lives in the bulk, due to the linear CS coupling it can
directly affect the spectrum of open string modes on the brane. 

There is no covariant action that produces the self-duality equation (\ref{sd}) as its equation
of motion\cite{Marcus:1982yu}. However, we can think of the self-dual theory as the chiral half
of the non-chiral theory defined by 
\be
\label{lten}
{\cal L}_{\rm 10-d} \, = \, 
\textstyle{1\over 2} | F_{\it 5} |^2 + d\spc C_{\it 4} \wwwedge F_{\it 5}\,
\ee
were the fields $F_{\it 5}$ and $C_{\it 4}$ (initially) need to be viewed as independent variables.
The $C_{\it 4}$ equation of motion dictates that $dF_{\it 5} = 0$, which is the Bianchi identity
that tells us that $F_5$ is  the exterior derivative of a 4-form. The self-duality projection
is implement by the constraint that $F_{\it 5}=dC_{\it 4}$.

To be consistent with self-duality, we need to slightly rewrite the CS interaction term as 
\be
\label{cfint2}
\cL_{{\mathsf {CS}}} = 
C_{\it 4} \wwwedge \cF + \cA \wwwedge F_{\it 5} ,
\ee
where $\cF = d \cA$ is the abelian field strength. 
The interaction term (\ref{cfint2}) adds source terms to the
equations of motion for $C_{\it 4}$ and $F_{\it 5}$ (here $\delta_{\cal M}$ is the $\delta$-function localized on
the D5 world-volume~${\cal M}$) 
\ba
\label{newpot}
d F_{\it 5} = \cF \wwwedge \delta_{{}_{{\cal M}}} 
\, ,
 \, \qquad \qquad
\hstar F_{\it 5} - d \wwedge C_{\it 4}  =
 \,  \cA \wwwedge  \delta_{{}_{{\cal M}}}\, . 
 \ea
These equations are clearly consistent with the self-duality constraint $F_{\it 5}=\hstar F_{\it 5} $.

The field strength $F_{\it 5}$ is gauge invariant under $U(1)$ gauge transformations. However, we
see that due to the explicit appearance of the abelian gauge potential $\cA$, the interaction term (\ref{cfint2}) and resulting equations
(\ref{newpot}) are gauge invariant only under the combined gauge transformation
\ba
\label{gct}
\cA \to \cA + d \lambda\, , 
\ & & \nonumber \\[-3mm] 
& & \qquad \qquad d\wwwedge (d\eta  + \, \lambda\,  \delta_{{\cal M}} )
\, = 0 .
\\[-3mm]
C_{\it 4} \to C_{\it 4} + d\eta\, , & &  \nonumber
\ea
This demonstrates that the RR 4-form $C_{\it 4}$ contains the longitudinal mode of $\cA$, which is
a first indication that the linear CS-coupling is capable of producing a mass term for $\cA$.

To isolate the longitudinal mode, 
consider the 4-cycle $\bbeta$ with the Calabi-Yau manifold
that is dual to the 2-cycle~$\aalpha$. 
The $\bbeta$-period of $C_{\it 4}$ defines a 4-d scalar 
\be
\label{vphi}
\varphi = \int_{\bbeta} \! C_4
 \, . 
\ee
Using that $\bbeta$ intersects the D5 world volume
$\cal M$ at a single point, we deduce 
that $\varphi$ transforms under $U(1)$ gauge rotations as
\be
\varphi  \to \varphi - \lambda\, . 
\ee
The 4-d action for $\varphi$ thus necessarily involves the gauge invariant combination $d\varphi + \cA$. 

Upon KK reduction, 
the RR 4-form field $C_4$ gives rise to a massless 2-form field 
for every 2-cycle 
within the CY manifold and  a massless scalar
mode 
for every 4-cycle.
To simplify our formulas, let us assume that
there is only one such dual pair, the 2-cycle $\aalpha$ wrapped by the single  D5 and its dual 4-cycle~$\bbeta$.
The associated massless 2-form field is obtained via 
\be
\label{phis}
\ccc = \int_{\aalpha} C_4\, \, .
\ee
$\ccc$ is  
related to the scalar $\varphi$ in (\ref{vphi})
via the 10-d self-duality equation $F_{\it 5} = \hstar  F_{\it 5}$. To write this relation, we
introduce the unique
harmonic 2-form $\omega$ and 4-form $\beta$ on the CY, normalized such that
\be
 \int_{\aalpha} \omega \;  = 1 \,, \qquad \qquad \int_\bbeta \beta = 1\, ,
 \qquad \quad
 \int_{{}_{CY}}\! \!\omega \wwwedge \beta\; =\; 1 \, .
\ee
The two forms are related via Hodge duality 
\be
\label{hodge}
\hhstar_{{}^6} \, \beta = \mu^2\, \omega 
\, ,  \qquad \qquad  \int_{{}_{CY}} \!\! \omega \wwwedge \hstar_{{}^6} \, \omega
= {\mbox{\small 1}\over \mbox{\small $\mu^2$}}\, .
\ee
Now, to perform the KK-reduction, we expand
\be
F_{\it 5} \, = \, d\hspace{1pt}  \ccc\wwwedge \omega \; +\; (d\varphi +\cA)\wwwedge\beta\; +\,  \ldots
\ee
We read off that $F_{\it 5}$ is self-dual, provided we identify
\be
\label{cphi}
  \hhstar_{{}^{ 4}} d\hspace{1pt} \ccc\, = \, 
\mu^2\, (d \varphi + \cA)\, .
\ee

Let us now return to our phenomenological set-up with a D5 wrapping a 2-cycle
$\aalpha_{\rm V}$ in the visible region and a D5 wrapping a 2-cycle $\aalpha_{\rm H}$ in the hidden
region. In case $\aalpha_{\rm V}$ and $\aalpha_{\rm H}$ were both
 distinct homology cycles within the CY, then after KK-reduction, the two world volume gauge fields $\cA_H$ and $\cA_H$
would couple to distinct 2-form modes $\ccc_V$ and $\ccc_H$. This is not our
situation of interest. Instead, we will arrange things such that the KK reduction produces a single 2-form mode
$\ccc$, that couples to a linear combination of $\cA_V$ and $\cA_H$. This is naturally achieved as follows.

Although the hidden and visible regions  may be located on opposite sides
of the CY manifold, it is possible to choose the CY geometry such that the 2-cycles $\aalpha_V$ and $\aalpha_{\rm H}$ are in fact topologically
the same 2-cycle, so that, as elements in $H_2(CY)$,
\be
\label{hom}
\aalpha_{\rm V} = \aalpha_{\rm H}\, .
\ee
This equation 
states that a D5 wrapping $\aalpha_{\rm V}$ can be continuously deformed into a D5
wrapping $\aalpha_{\rm H}$. During this deformation, the D5 would first need to expand while moving
away from the visible region, then traverse the CY and finally contract
to wrap a small 2-cycle near  the hidden region. So each brane locally minimizes its energy.
 This situation is schematically depicted in fig~4.

\begin{figure}[hbtp]
\begin{center}
 \includegraphics[width=2.6in]{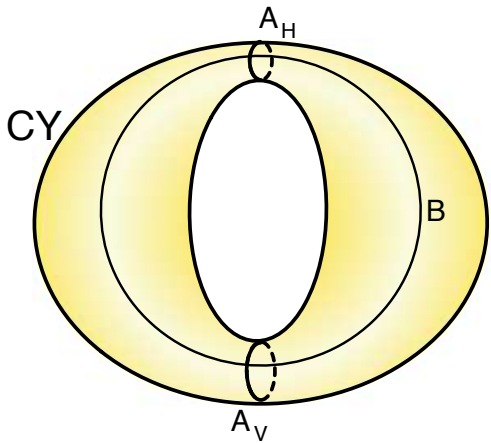} 
\end{center}
\vspace{-5mm}
\caption{A schematic depiction of our topological set-up. The visible and hidden D-branes wrap 
homologous cycles $\aalpha_{\rm V}$ and $\aalpha_{\rm H}$ within
the CY manifold, that share a common dual cycle $\bbeta$.  }
\end{figure}

The two homologous 2-cycles $\aalpha_{\rm V}$ and $\aalpha_{\rm H}$ share the same dual harmonic 2-form $\oomega$, and the same dual 4-cycle $\bbeta$ and associated harmonic 4-form $\beta$. We again
make the simplifying assumption that these are the only non-trivial 2- and 4-cycles around. The KK reduction of the 4-form $C_{\it 4}$ then proceeds as outlined above, except
that we must include the CS-coupling to both the visible and hidden branes. As a result, the 4-d duality relation
(\ref{cphi}) gets modified to
\be
\label{ncphi}
  \hhstar_{{}^{ 4}} d\hspace{1pt} \ccc\, = \, 
\mu^2\, (d \varphi + \cA_V + \cA_H)\, .
\ee
The rest of the story now proceeds as summarized in the introduction: the KK reduction yields a 4-d action for the
2-form field $\ccc$ of the form (\ref{een}), where $\mu^2$ is defined in (\ref{hodge}). The above relation 
(\ref{ncphi}) arises as the solution to equation of motion of the dual action
(\ref{twee}). Eliminating $H$ from the action (\ref{twee}) results in the St\"uckelberg mass term
for $\cA_V+\cA_H$.  

More insight is obtained by considering the superfield version of the story \cite{ALE}. The field $\varphi$
arises as the imaginary part of a complex superfield $T$, one of the K\"ahler moduli that 
parametrizes the size of the compact CY manifold. The real bosonic part of $T$ equals the volume
(in the Einstein frame) of the 4-cycle $\bbeta$
$$
T = \phi + i \varphi\, =\int_\bbeta (\sqrt{g} + i C_4)\, .
$$
The 4-d action of  $T$ is obtained via KK reduction of the IIB supergravity action.
Supersymmetry and gauge invariance dictate that its kinetic term takes the generic form
\be
\label{kahler}
\int \! d^4\theta\, K(T+\overline{T}+V_V+ V_H)
\ee
where $V_V$ and $V_H$ denote the vector multiplets on the visible and hidden D5-brane.  Upon performing the superfield integral, $K$ 
yields for the bosonic kinetic term
\ba
\label{smass}
K^{\prime\prime}(\phi) \bigl(  |d \varphi + \cA_V+\cA_H|^2 + |d\phi|^2\bigr)+{K^\prime}(\phi)( {\cal D}_V + {\cal D}_H) \, .
\ea
This includes a mass term for $\cA_+ = \cA_V + \cA_H$.  This mass in essence arises via a conventional Higgs mechanism,
with $\Phi= e^{-T}$ playing the role of the charged Higgs superfield.
\footnote{The complex superfield $T$  is equal to the action of a euclidean D3-brane instanton  that wraps the 4-cycle $\bbeta$; contributions to the effective action involving the Higgs 
field $\Phi = e^{-T}$  can be viewed as the contributions due to D3-brane instantons.}
In a complete setting, the real modulus 
field $\phi$  is -- via a modulus stabilzation mechanism driven by the coupling to other string 
scale degrees of freedom -- stabilized at some non-vanishing
expectation value $\avg{\phi}$. 
This translates into a finite vacuum condensate for the charged Higgs
superfield $\Phi$, that spontaneously 
breaks  the $U(1)$ gauge symmetry and produces the mass term for $\cA_+$

Eqn (\ref{smass}) indicates that the coupling to the modulus $\avg{\phi}$ may
also induce  an FI-term for
the hidden and visible brane theory. If there is $U(1)$ charged chiral
matter present on either the visible or hidden brane,  the D-term
constraint then enforces that some of these charged scalar fields 
must develop a vacuum expectation value. 
This is another mechanism for giving mass to the $U(1)$ bosons,
distinct from the bulk mechanism outlined above. For U(1) mediation,
we need to assume that these D-term driven masses are small compared
to the SUSY breaking scale, possibly by making certain choices of the
content of the hidden sector.
\bigskip
\bigskip

\newsubsection{Low energy parameters}

In this section we will discuss how SUSY breaking may get mediated
from the hidden to the visible 
brane, from the low energy point of view. Our discussion will be
brief, because after decoupling 
the heavy $U(1)$ gauge boson $\cA_+$, the mechanism  simply reduces to
$U(1)$ mediation via the  
light gauge boson $\cA_-$.

The gauge couplings of D-brane gauge fields are governed by closed string moduli
that paramerize the size of the cycles wrapped by the D-branes, and by
the periods of the NS 2 form $B$. The kinetic term of the two $U(1)$ gauge fields take the form
\ba
\label{kinetic}
{\cal W}  \; = \sum_{i=H,V}\, \frac{f_i(S_i, X_i)}{4}\, W_i^\alpha W_{i,\alpha} 
+ h.c.
\ea 
where $S_i$, $i=$ {\small  $V,H$} are closed string moduli fields, and $X_i$ $i=$ {\small $V,H$}
other open string matter fields,  that may contribute to supersymmetry breaking and mediation.  The gauge-coupling is given by  
The moduli $S_i$ are stabilized at some non-zero expectation
value. The $U(1)$ gauge couplings are then given by
\ba
\qquad \qquad g_i^2 = \frac{1}{{\rm Re} f_i(\langle S_i\rangle)}\, , 
\qquad  \quad \mbox{{\small$i= V,H$}}.
\ea
Hence, after canonically normalizing the gauge-fields, the heavy linear combination reads
\ba
\cA_+  \is \frac{1}{\sqrt{g_V^2 + g_H^2}} \left( g_V\cA_V + g_H\cA_H \right) \, . \\[1mm]
\cA^\prime \is \frac{1}{\sqrt{g_V^2 + g_H^2}} \left( g_H\cA_V - g_V\cA_H \right) \, .
\ea
The heavy field $\cA_+$ has a mass $m_+ = \mu  \sqrt{g_V^2+g_H^2}$ with $\mu$ defined in eqn (\ref{hodge}).  
The expression for $\mu$ in terms of geometric shape of the Calabi-Yau geometry depends on details of 
the supergravity reduction, that we will not attempt to evaluate here. It is clear, however, 
that for string size compactifications, the mass is of order the
string scale.\footnote{This conclusion is not dependent on whether the
  D-branes are localized at the bottom of a warped throat. 
The reason is that the integral (\ref{hodge}) is not a localized 5-d
mass term.} The $\cA_+$ gaugino marries off to the fermionic component
of $T$ to form a Dirac fermion with a mass of the same magnitude. We
will therefore assume that the heavy vector mutliplet decouples from
the low energy physics.\footnote{In scenarios with low string scale,
  the heavy field may still be relevant, as discussed in
  \cite{Ghilencea:2002da} .} The low energy spectrum thus only
contains the other linear combination $\cA_-$ 
and the corresponding light gaugino.

The $U(1)$ gaugino can receive a mass via gravity mediation or gauge mediation, or a combination of the two. In both cases, we imagine that the SUSY breaking contributions originate in the hidden region,
and first communicate their effect to the hidden sector brane.
Let $F_S$ denote the F-term contribution to the hidden closed string modulus $S_H$, 
and  $F_X$ the F-term of the hidden matter field  $X_H$, generated via 
possible gauge mediated contributions to supersymmetry breaking. Then the 
$U(1)$ gaugino mass is  (up to
corrections suppressed by the heavy linear combination's mass), 
\ba
\label{eqn:gauginoMass}
M_{\lambda^{\prime}} = \frac{g_V^2}{g_V^2 + g_H^2}\, (F_S \partial_S f_H + F_X \partial_X f_H)
\ea

Summarizing, the parameters relevant for the low energy theory are the
gauge-coupling of the light linear combination, 
\be 
g^\prime =\frac{ g_Hg_V}{ \sqrt{g_H^2+g_V^2}},
\ee 
 and the corresponding gaugino mass eqn (\ref{eqn:gauginoMass}). In
 addition, there can be a  
 D-term, as read off from eqn (\ref{smass}). If the D-term is small
 compared with the gaugino soft mass then one is lead to the model
 recently considered in \cite{LPWY}. If more structure is
 added then models such as the one considered in \cite{Kaplan:1998jk}
 can be constructed.

In the set-up as discussed so far, we have assumed that the hidden and visible gauge bosons
are associated with non-anomalous gauge symmetries.
In principle, it is also possible to consider the case when the hidden and visible $U(1)$ are both anomalous. The RR-form mechanism for generating a $U(1)$ masses is indeed 
a familiar part of  the Green-Schwarz anomaly cancelation mechanism. Models of 
SUSY breaking and mediation with   anomalous
$U(1)$'s has been previously studied in \cite{Binetruy:1996uv,Dvali:1996rj}. 

\bigskip

\bigskip

\newsection{Summary and Discussion}

We have shown that string phenomenological scenarios with visible and hidden sector
localized on D-branes, the coupling of the D-brane gauge theory to RR p-forms in the bulk 
gives rise to a robust SUSY breaking mediation mechanism, that remains effective when 
the two sectors  are separated by a large distance compared to the string length. Our exposition
specialized to $U(1)$ gauge theories on D5-branes in IIB string theory, but the same mechanism  
works in type IIA and heterotic string theory. In IIA constructions, the gauge theory
typically lives on D6-branes, which have a linear CS coupling $C_5 \wwwedge F$ to an
RR 5-form. In case the visible and hidden sector D6-branes wrap homologous
3-cycles, the RR coupling produces a mass-term for the diagonal combination of the two
vector bosons, leading to a low energy action of the form (\ref{vihi}) suitable for
$U(1)$ mediation. In the Appendix, we outline how $U(1)$ mediation is naturally
realized in the heterotic string.

We have not been specific about how supersymmetry may be broken on the hidden
brane. There are several possibiliities one could contemplate. Which of these  
is most adequate may depend  on the required scale of SUSY breaking,
and on whether $U(1)$ mediation  acts in combination with other 
mechanisms or is the main mechanism responsible for generating
the soft parameters.

In pure $U(1)$ mediation, one only needs a relatively low SUSY breaking
scale, but one needs the ability to tune this scale to a high accuracy \cite{LPWY}.  As a 
possible realization, we can look for a gauge theory on the hidden sector brane 
that spontaneously breaks SUSY. The D-brane technology 
for engineering such gauge theories 
has substantially improved over the last year, and a growing number 
of examples have been found \cite{nonsusybranes}. 
We can now imagine placing one of these SUSY breaking
D-brane systems at the tip of a warped deformed conifold \cite{KS}\cite{GKP}.  In this way, one can
discretely tune the SUSY breaking scale in the hidden sector by appropriately 
adjusting the quantized 3-form fluxes that support the conifold. 

Alternatively, one can consider scenarios that combines $U(1)$ mediation 
with another mechanism, such as gauge mediation or anomaly mediation.
Like gauge mediation, $U(1)$ mediation works for relatively low SUSY 
breaking scales. Moreover, in D-brane models of gauge mediation,
the visible and hidden sector are living on nearby stacks of branes, one may 
easily encounter situations where both sectors have a common $U(1)$ 
factor.

$U(1)$ mediation may also mix with anomaly mediation.  The two form
a natural pair,  since both are flavor blind bulk mediation mechanisms that can
communicate SUSY breaking between geometrically sequestered sectors. 
It was recently shown that the sequestering, that is a necessary precondition 
for anomaly meditation, can be realized in string compactifications where the 
SUSY breaking takes place at the bottom of a warped conifold. In case a visible 
$U(1)$ gauge  coupling gets a contribution from a hidden sector located in the 
non-SUSY region, one can imagine that sequestering fails for this particular 
coupling only. The spectrum and phenomenology of this mixed scenario,
for the special case of hypercharge $U(1)_Y$ mediation, is studied in \cite{hypno}.

\bigskip
\bigskip

\noindent
{\large \bf Acknowledgements}
 
We acknowledge useful discussions with R. Dermisek,
P. Langacker, G. Paz, and C. Vafa.  This work was
supported by the National Science Foundation under grant PHY-0243680. Any opinions,
findings, and conclusions or recommendations expressed in  this material are
those of the authors and do not necessarily reflect the views of the National Science
Foundation.

\bigskip
\bigskip
\medskip

\renewcommand{\newsection}[1]{
\setcounter{section}{1} 
\setcounter{subsection}{0} \addcontentsline{toc}{section}{\protect
\numberline{\roman{section}}{{\rm #1}}} \vglue .0cm \pagebreak[3]
\noindent{\large \bf  #1}\nopagebreak[4]\par\vskip .1cm}
\renewcommand{\theequation}{A-\arabic{equation}}
\setcounter{equation}{1}

\newsection{Appendix: $U(1)$ mediation in the heterotic string}

Although our focus in this paper is on the type II setting with
D-branes, the mechanism we have discussed can also be implemented in
the heterotic string. 

Consider the $E_8 \times E_8$ heterotic string compactified on a
Calabi-Yau manifold. In order to get chiral matter we turn on
holonomy for the ten-dimensional gauge groups and solve the Dirac
equation for the ten-dimensional gaugino in this background. The low
energy gauge group is the commutant of the holonomy group in $E_8
\times E_8$. This gives rise to two four-dimensional sectors, one
sector descending from the first $E_8$ which will be our visible
sector, and the other from the second $E_8$ which will be the hidden
sector where SUSY breaking takes place.

If the holonomy group includes a $U(1)$ factor, i.e. there is a
non-trivial line bundle $L$ with first Chern class $c_1(L)$, then
this $U(1)$ is also part of the unbroken gauge group since it
commutes with the holonomy. However the 10-d action has
couplings of the form
\be\label{hetStuck} {\mathcal L}_{10} \quad \supset \quad  (dB -
\omega_L - \omega_R)^2 \ee
where $\omega_{L,R}$ are the Chern-Simons three-forms for the two
$E_8$ gauge groups. Upon KK reduction of the B-field along the
harmonic form in the class $c_1(L)$ we get a non-universal axion
with a St\"uckelberg coupling to our $U(1)$ gauge field. Thus the
unbroken $U(1)$ generically gets a large mass and is removed from
the low energy spectrum.

However, analogous to the type II set-up, we may also turn on a line
bundle $L'$ in the hidden $E_8$ with first Chern class $[c_1(L)] =
[c_1(L')]$, which yields a corresponding unbroken $U(1)$ in the
hidden sector. The ten-dimensional couplings (\ref{hetStuck}) give
rise to
\be {\mathcal L}_{4} \quad \supset \quad (\del_\mu a - A_\mu^L -
A_\mu^R)^2 \ee
in four dimensions. Thus we see that the linear combination $A_\mu^L
+ A_\mu^R$ swallows the axion $a$ and gets a large mass, and we are
left with the linear combination $A_\mu^L - A_\mu^R$ which couples
to charged matter both in the hidden and the visible sector. We can
still give this extra $Z'$ a mass by a conventional Higgsing in the
hidden sector.

This method for generating massless $U(1)$'s has been known since
the early days of heterotic model building (see eg.
\cite{Witten:1985bz}, where the role of the massless $U(1)$ was
played by hypercharge). However it does not seem to have been
considered in the context of mediation of SUSY breaking. For more
recent work on heterotic model building with $U(1)$ factors, see
\cite{Blumenhagen:2005ga}.

\bibliographystyle{jhep}
\end{document}